\documentclass[prl,aps,twocolumn,english,superscriptaddress, nofootinbib]{revtex4}

\usepackage[]{fontenc}
\usepackage[latin9]{inputenc}
\usepackage{textcomp}
\usepackage{amsmath}
\usepackage{graphicx}
\usepackage{amssymb}
\usepackage{color}
\usepackage{soul}

\makeatletter

\pdfpageheight\paperheight
\pdfpagewidth\paperwidth

\newcommand{\lyxmathsym}[1]{\ifmmode\begingroup\def\b@ld{bold}
  \text{\ifx\math@version\b@ld\bfseries\fi#1}\endgroup\else#1\fi}

\@ifundefined{textcolor}{}
{%
 \definecolor{BLACK}{gray}{0}
 \definecolor{WHITE}{gray}{1}
 \definecolor{RED}{rgb}{1,0,0}
 \definecolor{GREEN}{rgb}{0,1,0}
 \definecolor{BLUE}{rgb}{0,0,1}
 \definecolor{CYAN}{cmyk}{1,0,0,0}
 \definecolor{MAGENTA}{cmyk}{0,1,0,0}
 \definecolor{YELLOW}{cmyk}{0,0,1,0}
 }
\newcommand{\urusi}{$\mathrm{URu_2Si_2}$}
\newcommand{\xrusi}{$\mathrm{XRu_2Si_2}$}

\newcommand{\tk}{$\mathrm{T_K}$}
\newcommand{\tho}{$\mathrm{T_{HO}}$}

\begin{document}

\title{Formation of the coherent heavy fermion liquid at the `hidden order' transition in URu$_2$Si$_2$}

\author{Shouvik Chatterjee \footnote{These authors contributed equally to this work.}}

\affiliation{Laboratory of Atomic and Solid State Physics, Department of Physics,
Cornell University, Ithaca, New York 14853, USA}

\author{Jan Trinckauf \footnotemark[\value{footnote}]}

\affiliation{Leibniz Institute for Solid State and Materials Research IFW Dresden, Helmholtzstra\ss e 20, 01069 Dresden, Germany}

\author{Torben~H\"anke}

\affiliation{Leibniz Institute for Solid State and Materials Research IFW Dresden, Helmholtzstra\ss e 20, 01069 Dresden, Germany}

\author{Daniel E. Shai}

\author{John W. Harter}

\affiliation{Laboratory of Atomic and Solid State Physics, Department of Physics,
Cornell University, Ithaca, New York 14853, USA}

\author{Travis J. Williams}

\author{Graeme M. Luke}

\affiliation{Dept. of Physics and Astronomy, McMaster University, 1280 Main St. West, Hamilton, Ontario L8S 4M1, Canada }

\author{Kyle M. Shen}

\affiliation{Laboratory of Atomic and Solid State Physics, Department of Physics,
Cornell University, Ithaca, New York 14853, USA}

\affiliation{Kavli Institute at Cornell for Nanoscale Science, Ithaca, New York
14853, USA}

\author{Jochen Geck}

\email[email: ]{j.geck@ifw-dresden.de}

\affiliation{Leibniz Institute for Solid State and Materials Research IFW Dresden, Helmholtzstra\ss e 20, 01069 Dresden, Germany}

\date{\today}

\begin{abstract}
In this article we present high-resolution angle-resolved photoemission (ARPES) spectra of the heavy-fermion superconductor \urusi. Measurements as a function of both excitation energy and temperature allow us to disentangle a variety of spectral features, revealing the evolution of the low energy electronic structure across the hidden order transition. Already above the hidden order transition our measurements reveal the existence of weakly dispersive states below the Fermi level that exhibit a large scattering rate. Upon entering the hidden order phase, these states transform into a coherent heavy fermion liquid that hybridizes with the conduction bands. 
\end{abstract}

\maketitle

The interactions between localized and delocalized electrons in the so-called heavy fermion materials result in fascinating and unexpected quantum phenomena which continue to challenge condensed matter researchers. One of the most prominent examples is the enigmatic `hidden order' (HO) state in URu$_{2}$Si$_{2}$ which is characterized by a large loss of entropy at \tho=17.5\,K \cite{Paltra, Mydosh:2011go}. Although a multitude of theoretical scenarios have been proposed to explain the HO transition \cite{chandra, Dubi:2011cs,Elgazzar:2009cx,Haule:2009jz,Chandra:2012ub,Ikeda:2012jn}, our lack of understanding of the complex and still debated electronic structure of \urusi\/ remains the major obstacle to developing a definitive understanding of this phase\,\cite{Ito:1999wv, Denlinger:2001jy}. Here we disentangle the low-energy electronic structure of \urusi by means of angle-resolved photoemission spectroscopy (ARPES) as a function of both excitation photon energy and temperature. We directly observe that precisely at \tho\/ localized and fluctuating electronic states rapidly hybridize with light conduction states, forming a well-defined heavy band coincident with a dramatic reduction in the scattering rate. We thereby demonstrate that in \urusi , the formation of the coherent heavy fermion liquid occurs via a thermodynamic phase transition into the HO phase. This behavior is in stark contrast with the gradual crossover expected in Kondo lattice systems, suggesting the possibility of multiple pathways towards the creation of heavy fermionic states.

\begin{figure}[b]
\includegraphics[width=1\columnwidth]{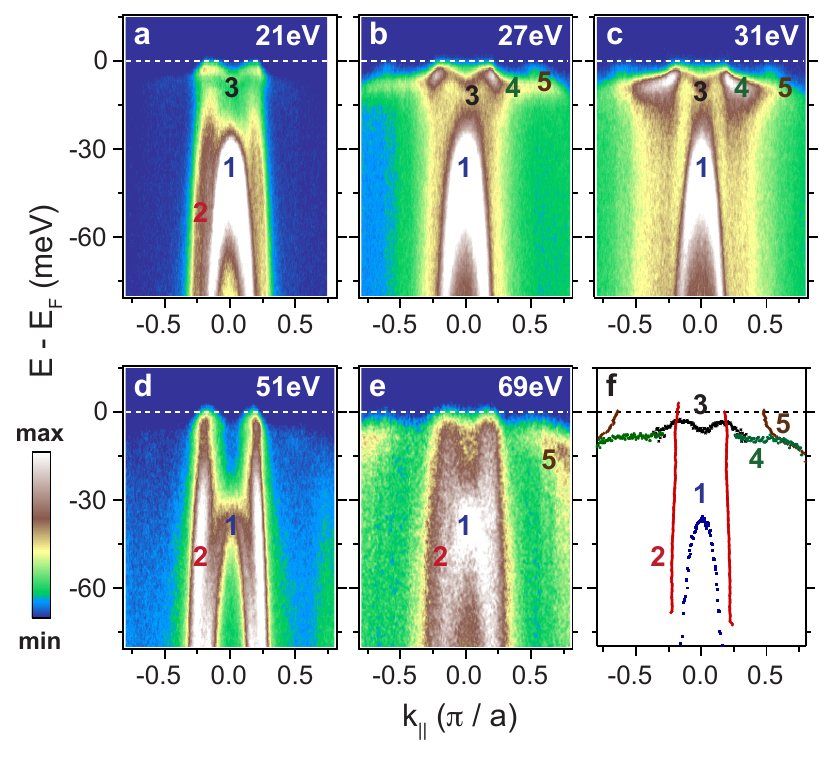}
\caption{\textbf{E-k maps at different excitation energy.} a-e, ARPES spectra along the $(0,0)$ to $(\pi,0)$ direction for different excitation energies (noted in the top right of each image) measured at 2~K, deep inside the hidden order (HO) phase. f,  Dispersions of all the different features obtained from fits to corresponding EDC/MDCs.} \label{edep}
\end{figure}

The present experiments were performed at the 1$^{3}$-endstation on beamline UE112PG2 at the Berlin Synchrotron BESSY II using a Gammadata R4000 anlyser with an overall energy resolution better than 7 meV and a base temperature lower than 2K.  Single crystals of \urusi\/ were cleaved along the $c$-axis \emph{in situ} at a base pressure of better than 4 $\times$ 10$^{-11}$ torr. Polarization of the incident photon beam was set at Linear Vertical (LV) unless mentioned otherwise. Fermi energy is determined by measuring a  polycrystalline gold film evaporated near the sample with a precision of better than 1 meV.

\begin{figure*}[t!]
\includegraphics[width=0.95\textwidth]{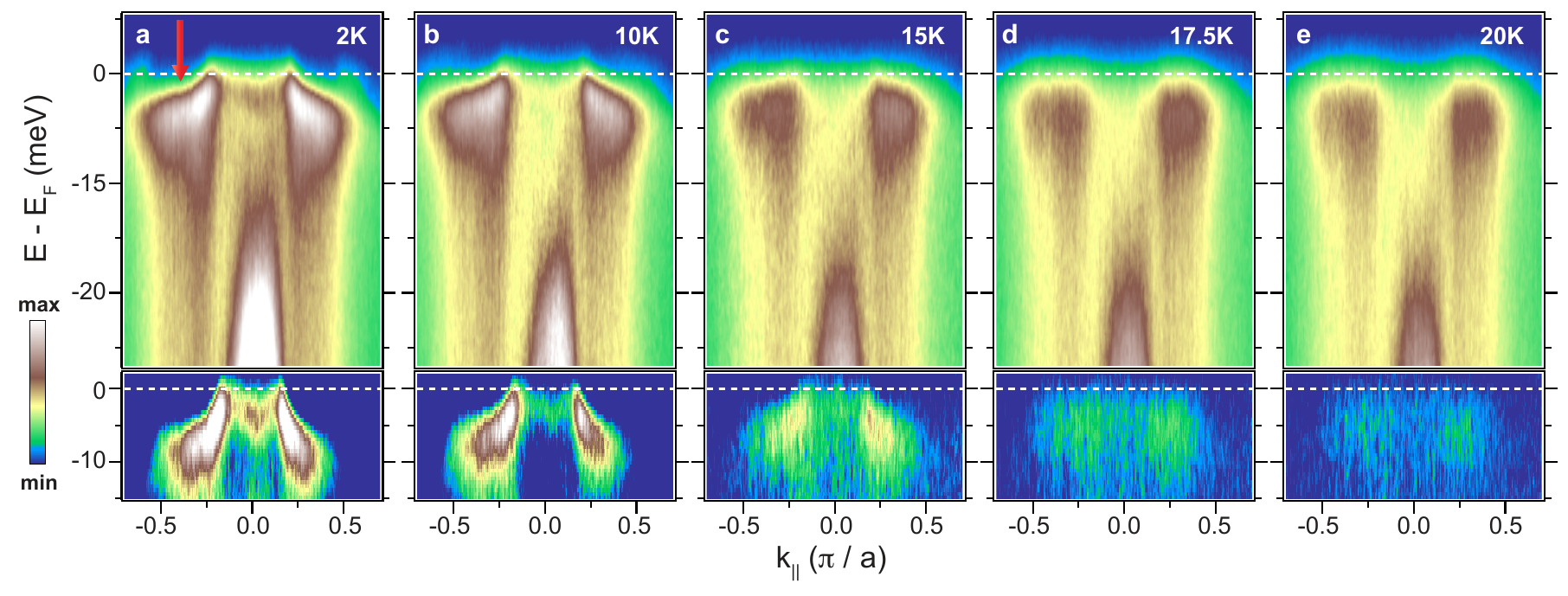}
\caption{\textbf{ Incoherent-coherent transition of the f-derived states at $\mathbf{T_{HO}}$.} a-e, Temperature evolution of the ARPES intensity plots of \urusi\/ measured along the (0,0) - ($\pi$,0) direction at 31 eV photon energy over the temperature range 2-20 K. In the lower panel ARPES spectral maps obtained after subtracting the corresponding intensity map at 25 K is shown. Color scale has been adjusted to show only the positive part of the subtracted spectrum. Note that all the spectral maps in the lower panel are plotted keeping the range of the color scale fixed. Below \tho\/ a coherent heavy fermionic band rapidly emerges which simultaneously becomes sharper and more dispersive as the sample is cooled down. The red arrow in a indicates the momentum at which the EDCs shown in Fig. \ref{fig:EDC} are taken.
\label{fig:Tdep}}
\end{figure*}

In Fig.\,\ref{edep} we show ARPES spectra along the (0,0) - ($\pi$,0) direction deep within the HO phase at a variety of different photon energies. The spectra in Fig.\,\ref{edep}\,a-e exhibit a dramatic dependence on the incident photon energies, revealing a multitude of electronic states near $(k_{x} = 0, k_{y} = 0)$, some of which have not been clearly delineated by previous photoemission studies. We emphasize that at no single photon energy are we able to clearly distinguish all five features, thus underscoring the importance of photon energy dependent measurements in revealing and disentangling the complete electronic structure of \urusi. A compilation of these different features is shown in Fig.\,\ref{edep}\,f. Feature 1 has been previously shown to be of surface origin, while feature 2 corresponds to a light hole-like band which has been attributed to a bulk state \cite{ Boariu:2010dj, SantanderSyro:2009ec}. Feature 3 exhibits an `M'-shaped dispersion also reported at 7 eV \cite{Yoshida:2010fg, Yoshida:2012ia}, and is connected to a relatively flat band (feature 4) ostensibly of predominantly $5f$ character. Finally, hole-like states (feature 5) that cross the Fermi level $E_{F}$ at $k_{x}$ $\approx\/$ 0.54~$\pi/a$ form propeller-shaped Fermi surface (FS) sheets, also observed in quantum oscillation measurements \cite{Ohkuni1999,Hassinger2010}. Another FS sheet reported by Shubnikov-de Haas (SdH) oscillations \cite {Shishido} exhibits an extremal $k_{F}$ similar to our feature 2, the light hole band. However, the SdH experiments also indicate that this FS sheet is closed along the (001) direction and only appears above a magnetic field of 21 T. At face value, this strong $k_{z}$ dependence appears inconsistent with our data, however, this could be resolved by the fact that our measurements are performed in the absence of a magnetic field.

By changing photon energy, we can probe different values of $k_{z}$ along the (001) direction and can therefore determine the electronic dispersion perpendicular to the Ru$_{2}$Si$_{2}$ planes. We do not observe any appreciable dispersion along $k_{z}$ for features 2, 3, and 4, while feature 1 has already been ascribed to a surface-derived origin and feature 5 is  apparent at only very few photon energies. The main effect of varying photon energies here is to strongly modulate the photoelectron matrix elements of these different features, suggesting that these states have substantially different orbital character.

We will concentrate primarily on features 2, 3 and 4 in Fig.\,\ref{edep}, all three of which undergo dramatic modifications across \tho. The lack of obvious $k_{z}$ dispersion makes it difficult to definitively assign these features to bulk states. Nevertheless, their strong temperature dependence allows us to state conclusively that they are tied to the onset of HO in the bulk. Moreover, the absence of feature 3 in Rh-doped samples where the HO state is destroyed \cite{ Yoshida:2010fg} further supports the assignment to bulk-derived states. Having identified the electronic states of interest, we now address their evolution across \tho. In what follows we will refer to the states corresponding to feature 3 (`M' shaped band) and feature 4 (flat band) as heavy fermion states and to feature 2 as the conduction band. 

\begin{figure}[t!]
\includegraphics[width=1\columnwidth]{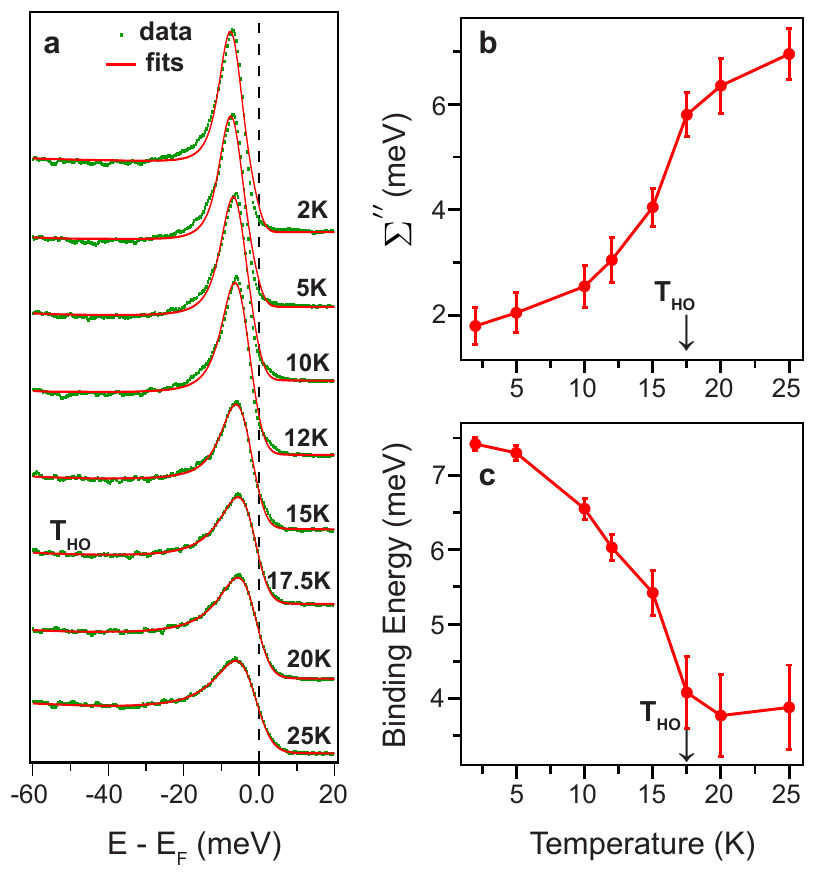}
\caption{\textbf{Suppression of quasiparticle scattering rate upon entering the hidden order phase. } a, Temperature dependence of the EDCs taken at the red arrow in Fig. \ref{fig:Tdep}\,a with corresponding fits (solid red lines). An instrumental resolution of 6 meV was used in the fits, as obtained from a gold reference. b,  Change in the imaginary part of the spectral function $\Sigma^{''}$ and c quasiparticle binding energy with temperature as extracted from fits in a. A sharp drop in magnitude observed  across \tho\/ shown  in \b indicates a dramatic enhancement of the lifetime of the quasiparticles on entering the hidden order phase.  
\label{fig:EDC}}
\end{figure}

To investigate the heavy fermion states, we set $h\nu=31$\,eV, a photon energy at which these states can be easily tracked. As shown in Fig.\,\ref{fig:Tdep}, above \tho\ only diffuse spectral weight is observed close to the Fermi level, indicating large scattering rates. As the temperature is lowered below \tho, a well-defined heavy fermion band forms, which becomes progressively sharper and more dispersive upon cooling. This development is even more apparent in the lower panels of Fig.\,\ref{fig:Tdep}, where the corresponding spectrum taken at 25\,K has been subtracted. In more conventional Kondo lattice systems, coherent heavy fermion bands develop only gradually below the Kondo temperature \tk, which is approximately 70 K for URu$_{2}$Si$_{2}$. In contrast, we observe only incoherent, localized states consistent with recent optical spectroscopy measurements \cite{Nagel}, which suddenly gain coherence below \tho\/.

To better quantify this temperature dependence, we have analyzed the energy distribution curves (EDCs) at the momentum indicated (red arrow) in  Fig.\,\ref{fig:Tdep}\,a. The data were fit to a lorentzian plus a temperature-independent Shirley background \cite{Shirley}, multiplied by a Fermi-Dirac function and finally convolved with the instrumental resolution. As can be observed in Fig.\,\ref{fig:EDC}\,b, the scattering rate obtained from the width of the lorentzian exhibits a sharp drop precisely at \tho. A similar temperature dependence has been observed in inelastic neutron scattering measurements, where the intensity of low energy spin excitations is greatly diminished upon entering the hidden order phase \cite{Wiebe:2007}. Moreover, a decrease in the electronic relaxation rate upon entering the HO phase has also been reported in recent pump-probe experiments \cite{ Dakovski:2011}.
The development of the dispersion is reflected in the shift of the peak of the EDC by approximately 4\,meV (Fig.\,\ref{fig:EDC}\,c), which is  consistent with optical spectroscopy \cite{Bonn, Hall}, transport \cite{Jeffries:2007, Mentink:1996} and tunneling measurements \cite{Aynajian:2010bz}. We note that this energy shift tracks the typical temperature dependence of an order parameter, supporting the notion that the observed changes in the electronic structure are directly related to the hidden order parameter. Indeed, this suggests that the changes in the electronic density of states at the HO transition which are often referred to as the hidden order gap are instead associated with the hybridization which gives rise to the heavy fermion states.

\begin{figure*}
\includegraphics[width=1\textwidth]{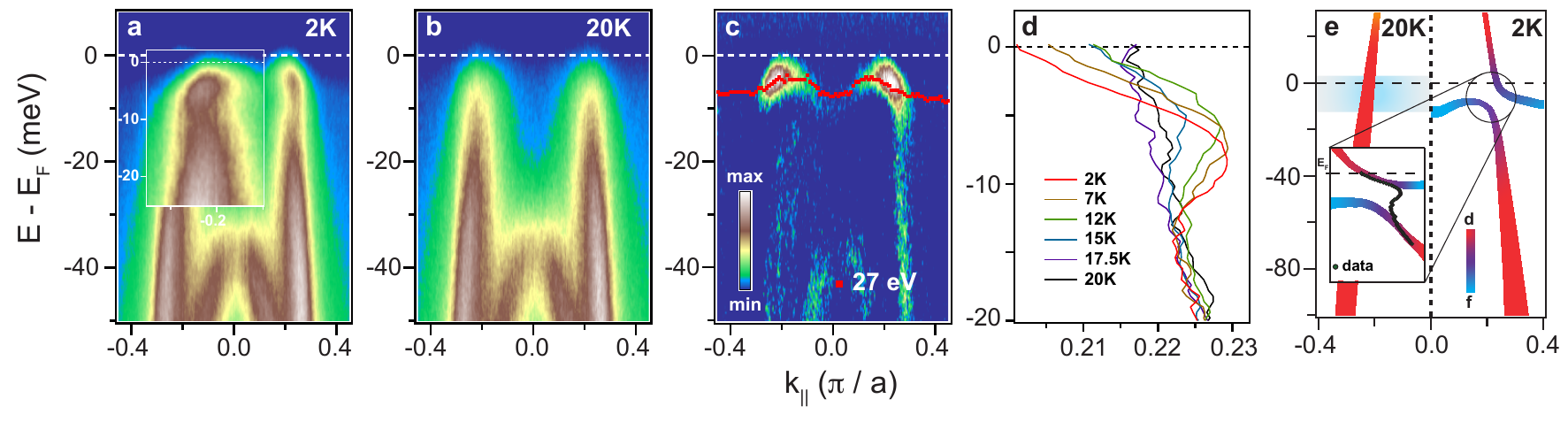}
\caption{\textbf{Rapid onset of hybridization at the hidden order transition.} a,  Angle resolved spectra along the $(0,0)$ to $(\pi,0)$ direction for 49 eV photon energy below a and above b  \tho\/.  Below \tho\/, a break appears in the dispersion, shown more clearly in the inset of a. c, Spectral map obtained after subtracting data in b from a clearly shows additional spectral weight appears below \tho\/ as the conduction band hybridizes with U 5f states. Dispersion of the `M' shaped feature as observed with 27 eV excitation energy (cf. Fig.\,\ref{edep} b)  is plotted on top of the subtracted spectrum showing the additional spectral weight follows exactly the dispersion of the `M' shaped feature.  d, Temperature dependence of the MDC derived dispersion of the conduction band. The kink feature progressively gets stronger and shifts towards higher binding energy as temperature is lowered below \tho\/. e, A schematic illustrating the changes in the electronic structure taking place across \tho\/. Inset shows how a kink appears in the dispersion as the bands develop mixed orbital character.\label{fig:kink}}
\end{figure*}

We now turn to the temperature dependence of the conduction band states across \tho. For this purpose we set $h\nu=49$\,eV, where the signal from the conduction band is strongly enhanced.
In Figs.\,\ref{fig:kink}\,a-b, we compare spectra measured at 2\,K and 20\,K, revealing very strong changes of the conduction band across \tho\/ due to the hybridization between the conduction band and the incoherent U $5f$ states as they develop coherence, an observation closely consistent with recent fourier-transform scanning tunneling spectroscopy measurements which track quasiparticle interference patterns \cite{Schmidt:2010bl, Yuan:2012bl}.
This is demonstrated clearly in Fig.\,\ref{fig:kink}\,c, where the difference of the spectra measured at 2\,K and 20\,K is presented. 
The additional spectral weight below \tho\/ tracks exactly the dispersion of the `M' band, showing that the formation of the coherent heavy fermion liquid goes hand in hand with the hybridization of the conduction band. This situation is summarized schematically in Fig.\,\ref{fig:kink}\,e, showing how variations in the photoelectron matrix elements due to rapidly changing orbital characters can give rise to an apparent dispersion anomaly as the bands hybridize.

Although the dispersion anomaly in Fig.\,\ref{fig:kink}\,a resembles a kink feature, we believe it is \emph{not} related to the coupling of the quasiparticles to a bosonic excitation. Apart from the arguments given above, there are a number of additional reasons why electron-boson coupling is unlikely to be responsible for the observed kink in the dispersion. First, the `kink' energy is characteristic of the boson energy, but is shown to be highly temperature dependent in Fig.\,\ref{fig:kink} d, vanishing above $T_{HO}$. Second, the ratio of  band velocity at higher binding energies to the the velocity at $E_{F}$ i.e  $v_{HBE} / v_{E_F}$ would be representative of the electron-boson coupling and mass renormalization, but the value of $v_{HBE} / v_{E_F} \approx $ 4.0 $\pm$ 0.2 at 2 K would signify an unphysically large value of the coupling strength, particularly for such a soft mode.

The emergence of the `M' feature observed here at $h\nu$ = 49, 27, and 21 eV agrees well with previous laser ARPES studies at $h\nu$ = 7\,eV \cite{Yoshida:2010fg,Yoshida:2012ia}, where it was interpreted in terms of a symmetry reduction and the resulting zone folding in the HO phase. However, the spectral weight arising from zone folding is typically much weaker than the original bands, whereas we observe that at certain photon energies (e.g. 49 eV) the `M' feature becomes as strong as the conduction band and the surface state. In addition, this feature coincides with the dispersion  of the coherent heavy fermion band below \tho. Our experiments therefore indicate that the emerging `M' feature is due to the formation of the heavy fermion liquid at \tho\ arising from the hybridization, and not from zone folding.

Our experiments reveal that in URu$_{2}$Si$_{2}$, the formation of the coherent heavy fermion liquid occurs via the thermodynamic phase transition into the HO state, in contrast to a gradual crossover below $T_{K}$, as is believed to be the case for other Kondo lattice systems. This indicates that there may exist multiple pathways to the formation of heavy fermionic quasiparticles, necessitating further studies of the electronic structure of additional $f$-electron systems. Furthermore, our measurements reveal that the putative hidden order gap observed in the electronic density of states by various probes is in fact associated with the hybridization which gives rise to the coherent heavy fermion quasiparticles. Our work reveals that the HO phenomenon is directly tied to the hybridization between the $5f$ states and the conduction band, an interaction which is blocked above \tho \/ and only becomes active inside the HO phase. Finally, the abrupt drop observed in the quasiparticle scattering rate through the HO transition suggests that the single particle electronic excitation spectrum is directly sensitive to fluctuations above \tho\ and indicative of an order-disorder transition. 


 

\section*{Acknowledgements}

We would like to thank E. Rienks for assistance with the measurements at the UE112PGM2 beamline at BESSY II. We also acknowledge enlightening conversations with K. Becker, P. Coleman, J.C.S. Davis, M. Hamidian, P.D.C. King, G. Kotliar, G.A. Sawatzky, S. Sykora, and J. van der Brink. This work was supported by the National Science Foundation through a CAREER award (DMR-0847385), the Materials Research Science and Engineering Centers program (DMR-1120296, the Cornell Center for Materials Research), and a Research Corporation Cottrell Scholars award (20025). J.T. and J.G. acknowledge the support by DFG through the Emmy-Noether program (Grant GE1647/2-1).  Work at McMaster was supported by NSERC and the Canadian Institute for Advanced Research.




\newpage



\end{document}